\documentclass[a4paper,preprint,superscriptaddress]{revtex4-1}
\pdfoutput=1

\usepackage{amsmath}
\usepackage[sort&compress]{natbib}
\usepackage{natbib}

\usepackage{psfrag}
\usepackage{graphicx} 
\usepackage{color} 
\usepackage{subfigure}
\usepackage[small, margin=50pt]{caption}
\usepackage[pdftex]{thumbpdf} 
\usepackage[pdftex,bookmarks=true,bookmarksnumbered=true,]{hyperref}
\usepackage{verbatim}

\hypersetup{
	pdfauthor = {C. Wesp, A. El, F. Reining, Z. Xu, I. Bouras, C. Greiner},
	pdftitle = {Calculation of shear viscosity using Green-Kubo relations within a parton cascade},
	pdfkeywords = {shear viscosity, Green-Kubo,pQCD, three particle
	interactions, inelastic interactions, transport, BAMPS, alpha s} }

\begin{document}

	\title{Calculation of shear viscosity using Green-Kubo relations within a parton cascade}

	\author{C.\ Wesp}
	\author{A.\ El}
	\author{F.\ Reining}
	\affiliation{Institut f\"ur Theoretische Physik, Johann Wolfgang
	Goethe-Universit\"at, Max-von-Laue-Str.\ 1, D-60438 Frankfurt am Main, Germany}

	\author{Z.\ Xu}
		\affiliation{Institut f\"ur Theoretische Physik, Johann Wolfgang
	Goethe-Universit\"at, Max-von-Laue-Str.\ 1, D-60438 Frankfurt am Main, Germany}
		\affiliation{Frankfurt Institute for Advanced Studies, Ruth-Moufang-Str.\
		1, D-60438 Frankfurt am Main, Germany}
	
	\author{I.\ Bouras}	
	\author{C.\ Greiner}
	\affiliation{Institut f\"ur Theoretische Physik, Johann Wolfgang
	Goethe-Universit\"at, Max-von-Laue-Str.\ 1, D-60438 Frankfurt am Main, Germany}

\begin{abstract}
The shear viscosity of a gluon gas is calculated using the Green-Kubo relation.
Time correlations of the energy-momentum tensor in thermal equilibrium are
extracted from microscopic simulations using a parton cascade solving various
Boltzmann collision processes. We find that the pQCD based gluon bremsstrahlung
described by Gunion-Bertsch processes significantly lowers the shear viscosity by a factor of $3 - 8$ compared to
 elastic scatterings. The shear viscosity scales with the coupling as $\eta \sim
 1/(\alpha_s^2 \ \log(1/\alpha_s) )$.  For constant $\alpha_s$ the shear viscosity to entropy
density ratio $\eta/s$ has no dependence on temperature. Replacing the
pQCD-based collision angle distribution of binary scatterings by an isotropic
form decreases the shear viscosity by a factor of $3$.
\end{abstract}


\date{\today}

\maketitle

\section{Introduction}
Results from the Relativistic Heavy Ion Collider (RHIC) at Brookhaven and the
Large Hadron Collider (LHC) at CERN suggest the formation of a new state of
matter, the quark gluon plasma (QGP) in ultra-relativistic heavy ion collisions.
Large values of the elliptic flow $v_2$ observed in these experiments
\citep{Ackermann:2000tr,PhysRevLett.91.182301,Adams:2003am,PhysRevLett.94.122303,
PhysRevLett.98.242302,PhysRevC.80.024909,lhc} lead to
the indication that the QGP behaves like a nearly perfect fluid, i.e. its shear
viscosity to entropy density ratio is small. This makes dissipative
hydrodynamics a promising candidate to describe the collective behavior of QGP
\citep{Song:2007ux,PhysRevC.78.034915,Teaney:2009qa,Schenke:2010nt,Song:2011qa,Schenke:2011tv}.
However, hydrodynamics only works if a system is close to thermal equilibrium,
which is not applicable for the state of matter shortly after the initial
heavy-ion collision where transport theory based models are a more suitable
approach \citep{PhysRevC.71.064901,Xu2008a,Ferini:2008he,Molnar:2001nk}. One of
today's challenge in theoretical nuclear physics is to find a transition between
the transport and hydrodynamic description as well as an appropriate set of
transport parameters, for example the shear viscosity $\eta$ of the collective
medium.  In \citep{1126-6708-2003-05-051} a full leading order pQCD evaluation
of shear viscosity in the collinear approximation of gluon splitting is
reported. Calculations with AdS/CFT correspondence in ${\cal N}=4$
super-symmetric Yang-Mills theory in large $N$ and strong coupling limit show a
small universal but non-vanishing lower bound of $\eta/s = 1/4 \pi$
\citep{Kovtun:2004de}. A recent analysis within a quasi-particle description and
employing a Boltzmann-Vlasov ansatz in the relaxation time approximation also
gives a small number when extrapolating in the non perturbative
regime\citep{Bluhm:2010qf}. Considering the fully microscopic transport
description BAMPS (Boltzmann Approach of Multi-Parton Scattering) while
employing binary elastic $gg \to gg$ and also inelastic bremsstrahlung $gg
\leftrightarrow ggg$ pQCD collisions, the shear viscosity was extracted by using
approaches motivated from first and second order hydrodynamics
\citep{PhysRevLett.100.172301,El:2008yy}. A small viscosity to entropy ratio
close to the Ads/CFT conjecture resulted, which is due to the incorporation of
the bremsstrahlung processes being more efficient because of larger momentum
deflection. Similar ideas have been given in \citep{Chen:2009sm, Chen:2010xk} by
also using pQCD elastic and inelastic matrix elements and modeling the
equilibrium deviations of the distribution function in consistency with kinetic
theory. Most recently, in \citep{Fuini:2010xz} the shear viscosity is extracted
also from a microscopic transport description employing using the Green-Kubo
relation and perturbative QCD matrix elements for elastic scattering.

In this work we will extract the shear viscosity of a gluon gas from microscopic
transport calculations with BAMPS employing also directly the Green-Kubo
relation. The Green-Kubo relation does not rely on model assumptions or the form
of equilibrium deviations. Thus this work can provide an independent cross check
to the previously published works \citep{PhysRevLett.100.172301,El:2008yy}. In
principle, the present calculations are rather close to those of \citep{Fuini:2010xz}. In Sec.
\ref{intro_1} we give a brief overview of the method. In Sec. \ref{sec:BAMPS}
the parton cascade BAMPS and numerical setups are briefly introduced. We
demonstrate in Sec. \ref{sec:methodology} how to extract the shear viscosity
numerically. The results are presented in Sec. \ref{numerical_results} and are
compared to previously published calculations. The dependence of the shear
viscosity on the pQCD coupling constant $\alpha_s$, temperature, and the
distribution of the collision angle are investigated and discussed. A comparison
with recent results presented in \cite{Fuini:2010xz} are be discussed in Sec.
\ref{angular_dependence}.

\section{Green-Kubo Relations}\label{intro_1}

Green and Kubo showed in serial papers \citep{kubo1957statistical2,kubo1957statistical}
that transport coefficients like heat conductivity, shear- and bulk viscosity
can be related to the correlation functions of the corresponding flux or 
tensor in thermal equilibrium. The physical motivation is given by Onsager's
regression hypothesis \citep{onsager}, stating that fluctuations are present
in every equilibrated system. Dissipation of fluctuations has the
same origin as the relaxation towards equilibrium once the system is
disturbed by an external force. Both the dissipation and the relaxation time
scales are determined by the same transport coefficients.

In this paper we concentrate on shear viscosity $\eta$. 
The corresponding Green-Kubo relation has the following form \citep{book2}:

\begin{equation} \label{green_kubo_definition} 
	\eta = \frac{1}{10 \text{T}} \int_{0}^{+ \infty} \mathrm{d}t \int_V
	\mathrm{d}^3 r \ \langle \pi^{ij}({\bf r},t) \pi^{ij}(0,0)
	\rangle \,,
\end{equation}
where $T$ is the temperature,
$\pi^{ij}$, $i,j=x,y,z$ are components of the shear stress tensor, and
$\langle \cdots \rangle$ denotes the ensemble average in thermal equilibrium.
The correlation between shear components at time $t=0$ and at $t$ 
is the sum over $i$ and $j$. Furthermore one can show that 
$\langle \pi^{ij}({\bf r},t) \pi^{ij}(0,0) \rangle=
10 \langle \pi^{xy}({\bf r},t) \pi^{xy}(0,0) \rangle$.

The Green-Kubo relation (\ref{green_kubo_definition}) is the long wave 
limit $k, \omega \to 0$ of the Fourier transform of the integrated 
correlation function \citep{book2}. Its derivation can be found elsewhere
\citep{book1, esterban_book,searles:9727}.

Within kinetic theory, the correlation of shear components can be computed
by employing transport simulations for a particle system embedded
in a static volume. Simulations using the hadron cascade model 
UrQMD \cite{Bass:1998ca,Bleicher:1999xi} and the parton cascade model 
PCM \cite{Bass:2002fh} have been performed to calculate the shear
viscosity of a hadron gas \cite{Demir:2008tr} and of a quark-gluon 
plasma \cite{Fuini:2010xz}.

In this paper we use the parton cascade BAMPS
\cite{PhysRevC.71.064901,PhysRevC.76.024911}
to calculate the shear viscosity of a gluon gas
including perturbation QCD based bremsstrahlung processes $gg\leftrightarrow
ggg$.

\section{The parton cascade BAMPS and numerical setups}\label{sec:BAMPS}

The parton cascade BAMPS solves the ultrarelativistic Boltzmann equation
\begin{equation}
p^\mu \partial_\mu f(x,p) = C(x,p)
\end{equation}
for on-shell particles. Their interactions are simulated via Monte Carlo 
techniques based on the stochastic interpretation of transition 
rates \cite{PhysRevC.71.064901}. 
Collision probabilities of two particles in a spatial cell of a volume
of $V_{\rm cell}$ and within a time step $\Delta t$ are
\begin{equation}\label{eq_BAMPS_collProb}
P_{22,23} = v_{ \rm rel} \frac{ \sigma_{22,23} }
{ N_{\rm test} } \frac{\Delta t}{ V_{\rm cell} } \,,
\end{equation}
where $\sigma_{22,23}$ are the total cross section for a $2\to 2$ and 
a $2\to 3$ collision, respectively. $v_{ \rm rel} = (p_1 + p_2)^2/(2 E_1
E_2)$ denotes the relative velocity of the two incoming particles with four
momenta $p_1, p_2$. The probability for the back reaction $3\to 2$ is
accordingly
\begin{equation}
\label{collProb32}
  P_{32} = \frac{1}{8 E_1 E_2 E_3} \frac{I_{32}}{N^2_{test}} \frac{\Delta
  t}{V_{\rm cell}^2}\,,
\end{equation}
where $I_{32}$ is a quantity corresponding to a cross section for $3 \to 2$
processes and is given by integrating over the final states in an interaction
process \cite{PhysRevC.71.064901}.

Between the collisions the particles propagate via free-streaming.
Simulations of the space-time evolution of the particles are performed in a
static box with periodic boundary conditions.

In order to ensure an accurate solution of the Boltzmann equation,
the volume of spatial cells has to be chosen as small as possible, which
obviously enhances statistical fluctuations in the simulation. To reduce
these fluctuations a test particle method \citep{PhysRevC.71.064901} is 
introduced: the particle number is artificially increased by a factor of
$N_{\rm test}$. Thus, the
collision probability has to be reduced by the same factor to keep 
the particle mean-free path independent of the value of $N_{\rm test}$.

For the present simulations, the length of cubic cells is set to be a factor of
$3$ smaller than the mean free path of particles, and the box length is
a factor of $8$ larger than the cell length. $N_{\rm test}$ is chosen
to have on average $12$ test particles in each cell.

However, the ratio of mean free path to box length does not influence
the volume averaged correlation function as long as particles itself stay
uncorrelated, which is the case when the total number of particles is large enough
 \citep{wesp-master}.

\section{Extraction of Shear Viscosity}\label{sec:methodology}

We calculate $\langle \pi^{xy}({\bf r},t) \pi^{xy}(0,0) \rangle$
numerically according to the Green-Kubo relation (\ref{green_kubo_definition}).
The shear component $\pi^{xy}$ is defined as
\begin{equation}\label{eq_EM-Tensor_1}
	\pi^{xy}({\bf r},t) = T^{xy}({\bf r},t)=
\int \frac{\mathrm{d}^3 {\bf p}}{(2\pi)^3 E} \ p^x p^y 
f({\bf r}, t; {\bf p})
\end{equation}

In the numerical simulation the volume averaged shear tensor is used

\begin{equation}\label{eq_EM-Tensor_bamps}
 \bar \pi^{x y}(t) = \frac{1}{V}\sum_{i=1}^N \frac{p^x_i p^y_i}{E_i} \,,
\end{equation}
where the sum is over all particles in the box at time $t$.
$V$ is the volume of the box. The correlation of $\bar \pi^{xy}$ is obtained
by time and ensemble average in the limit $T_t \to \infty $ 
\begin{equation} \label{definition_autocorrelation}
 \langle \bar \pi^{xy}(t) \ \bar \pi^{xy}(0) \rangle = 
 \left \langle \frac{1}{T_t} \int_0^{T_t} \bar \pi^{xy}(t+t') \bar \pi^{xy}(t')\
\mathrm{d}t' \right \rangle =
 \left \langle \frac{1}{N_t} \sum_{j=0}^{N_t-1} \bar \pi^{xy}(i\Delta t+j\Delta
 t) \bar \pi^{xy}(j \Delta t)\ \right \rangle \,,
\end{equation}
where $N_t=T_t/\Delta t$ and $i = t/\Delta t$.

With (\ref{eq_EM-Tensor_bamps}) and (\ref{definition_autocorrelation}) the
receive the following Green-Kubo relation:

\begin{equation}\label{Green-Kubo-averaged}
\eta  = \frac{V}{10 \ T} \int_0^\infty \mathrm{d}t \ \left \langle \bar
\pi^{xy}(t) \bar \pi^{xy}(0) \right \rangle
\end{equation} 

\begin{figure}[h]
\includegraphics[width=0.45 \textwidth]{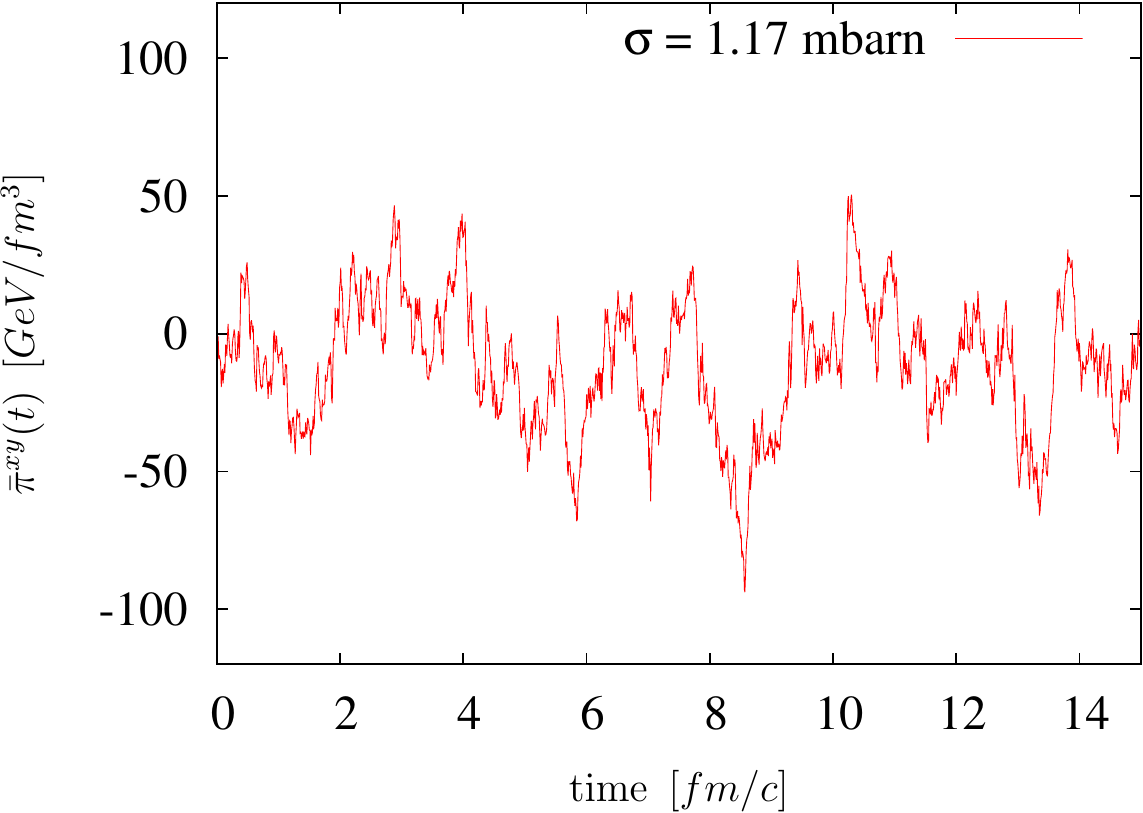}
\hspace{1cm}
\includegraphics[width=0.45 \textwidth]{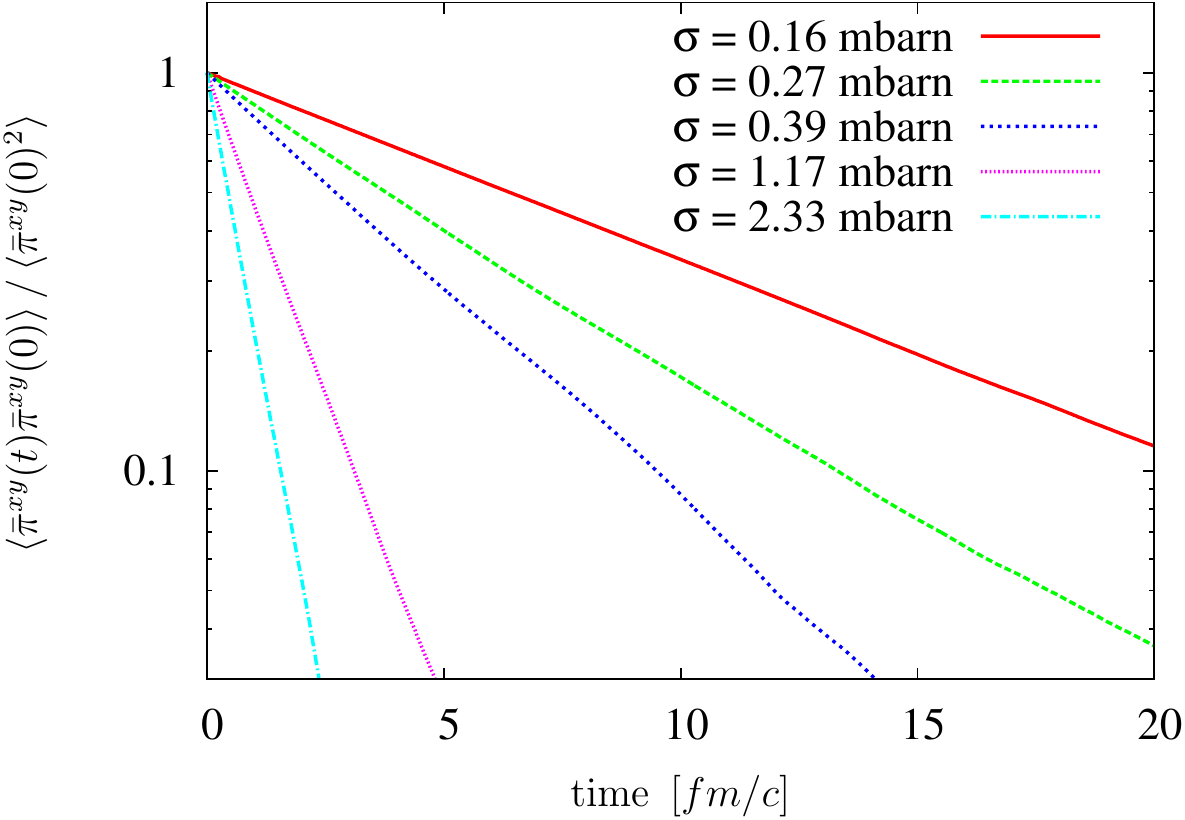}
\caption{(Color online) Left: Example for the equilibrium fluctuation of the
shear component $\pi^{xy}$. Right: Correlations for different isotropic and constant cross
sections. Results are normalised by $\langle (\bar \pi^{xy})^ 2 \rangle$.
}
\label{fig:Example}
\end{figure}

Figure \ref{fig:Example} shows the fluctuation of $\pi^{xy}$ in 
one event (left panel) and its correlations over time and ensemble average
(right panel) calculated using BAMPS.
Results are obtained for elastic binary collisions assuming constant
cross sections with isotropic distribution of the collision angle.
The initial distribution is the Boltzmann distribution at a temperature
of $T=400$ MeV
\begin{equation}
\label{boltzmann}
f({\bf p})= d_G \ e^{-E/T} \,,
\end{equation}
with the degeneracy factor $d_G=16$ for of gluons. We neglect Bose
enhancement, which is a quantum effect of gluons and has not been
implemented in BAMPS yet.

From Fig. \ref{fig:Example} we see that the correlation function is 
an exponentially decreasing function as expected for solutions of the
Boltzmann equation with stochastic interpreted cross-sections
\citep{reichl_book},
\begin{equation}
\label{relaxation}
 \langle \bar \pi^{xy}(t) \ \bar \pi^{xy}(0) \rangle = 
\langle \bar \pi^{xy}(0) \ \bar \pi^{xy}(0) \rangle e^{-t/\tau} \,,
\end{equation}
with a relaxation time scale $\tau$.
The initial variance of $\bar \pi^{xy}$ can be calculated analytically
and is given by
\begin{equation}\label{zerotime_autocorrelation}
\langle \bar \pi^{xy}(0) \ \bar \pi^{xy}(0) \rangle = 
\frac{4}{15} \frac{e T}{V} \,,
\end{equation}
where $e=3d_GT^4/\pi^2$ is the gluon energy density.
Inserting Eqs. (\ref{relaxation}) and (\ref{zerotime_autocorrelation})
into Eq. (\ref{green_kubo_definition}) we obtain
\begin{equation}\label{Green_Kubo_simple_1}
\eta =  \frac{4}{15} e \tau \,.
\end{equation}
Equation (\ref{Green_Kubo_simple_1}) is exactly the same as derived in Refs.
\cite{PhysRevLett.100.172301,Hirano:2005wx}, if one interprets the relaxation
time as the inverse of the transport collision rate. In general, the relaxation
time scale $\tau$ depends on temperature and individual matrix elements of
interactions. Its relationship with microscopic scales will be shown later
explicitly. The numerical task is now to find $\tau$ by an appropriate fit to
the numerically calculated correlation function.

\section{Numerical Results}\label{numerical_results}
In this section we present results of shear viscosity.
At first we assume constant cross sections for binary elastic collisions
with isotropic distribution of the collision angle. In this case we are able to
cross-check our numerical results with analytical calculations. 
Then we calculate the shear viscosity of gluons including pQCD-based 
elastic and bremsstrahlung processes.

\subsection{Isotropic cross sections}\label{iso}
For binary elastic collisions with isotropic distribution of the collision
angle the shear viscosity of an ultrarelativistic Maxwell-Botzmann gas is known
from the derivation in the Navier-Stokes approximation \citep{deGroot,Huovinen:2008te}, which is
\begin{equation} \label{analytic_shearvisco}
		 \eta^{NS} = 1.2654\ \frac{T}{\sigma_{22}} \,,
\end{equation}
where $\sigma_{22}$ is the total cross section for binary elastic collisions.
Comparing Eq. (\ref{analytic_shearvisco}) to Eq. (\ref{Green_Kubo_simple_1})
leads to $\tau=1.58/(n\sigma_{22})=1.58 \lambda_{mfp}$, i.e., in this case
the relaxation time is $1.58$ times the mean free path.

\begin{figure}[h] 
\includegraphics[width=0.8\textwidth]{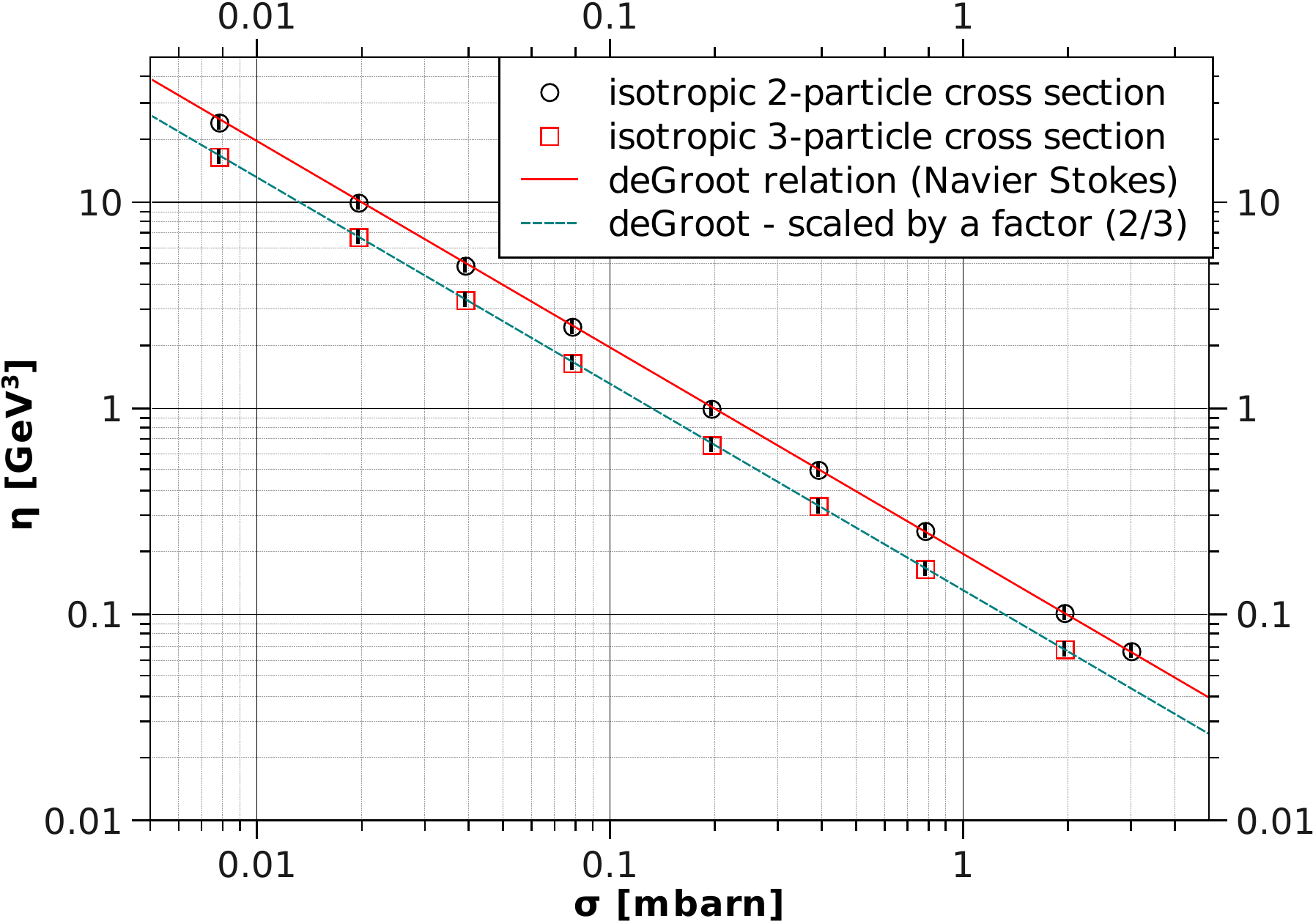} 
\caption{(Color online) Shear viscosity for a gas with constant and isotropic
cross sections. The open circles (squares) are results for elastic $2\to 2$ (inelastic
$2\leftrightarrow 3$) process. Temperature is fixed at
$T = 400 MeV$.}
\label{fig:eta_s_iso}
\end{figure}

Figure \ref{fig:eta_s_iso} shows our results in open circles, compared with
Eq. (\ref{analytic_shearvisco}) represented by a solid line. The standard
deviations are small and displayed within circle areas. We see an excellent agreement between
our numerical results and the analytical ones, which proves the applicability of
our numerical method within BAMPS.

We also calculate the shear viscosity for inelastic $2\leftrightarrow 3$
process with a constant cross section $\sigma_{23}$ and isotropic
distributions of the collision angles. In this case we 
have \citep{PhysRevC.71.064901}
\begin{equation}
\label{formel:I32}
  I_{32} = \frac{192}{d_G} \pi^2 \sigma_{23}
\end{equation}
for computing the interaction probability $P_{32}$ for the $3 \to 2$ process.

In Fig. \ref{fig:eta_s_iso} the results of the shear viscosity
for such inelastic processes, $\eta_{23}$, are given by the open squares.
Compared with the shear viscosity for binary elastic collisions, $\eta_{22}$, we find
that $\eta_{23}$ is a factor of $1.5$ smaller than $\eta_{22}$ for
$\sigma_{22} = \sigma_{23}$, as indicated
by the dashed line. This finding agrees with previously reported numerical
\citep{El:2010mt} and recent analytical calculation \citep{Lauciello_paper}.

\subsection{pQCD based cross sections}\label{result:pQCD}

We now consider elastic and bremsstrahlung processes of gluons based
on perturbation QCD (pQCD). The cross sections and matrix elements are the same
as used in previous studies 
\citep{PhysRevC.71.064901,PhysRevLett.100.172301,El:2008yy}.

For elastic interactions of gluons, $gg \leftrightarrow gg$, we use the Debye
screened cross section in small angle approximation

\begin{equation}
\label{csgg}
\frac{d\sigma_{gg\to gg}}{dq_\perp^2} \approx \frac{9\pi\alpha_s^2}{2}
\frac{1}{\left ( q_\perp^2+m_D^2 \right )^2} \,.
\end{equation}

The Debye screening mass is computed dynamically from the local particle
distribution $ f = f(p,x,t)$ via 

\begin{equation}
\label{mass_debyse}
m^2_D = d_G \pi \alpha_s \int \frac{\mathrm{d}^3 p}{(2\pi)^2} \frac{1}{p} \ N_c
\ f \,,
\end{equation}
where $d_G = 16$ is the gluon degeneracy factor for $N_c = 3$.

Inelastic $gg \leftrightarrow ggg$ processes are treated via an effective
matrix element based on the work by Gunion and Bertsch \citep{Gunion:1981qs}. Detailed
balance between gluon multiplication and annihilation processes is ensured by
the relation $\left |{\cal M}_{ gg \to ggg } \right |^2 = d_G \left | {\cal
M}_{ggg \to gg} \right |^2$. For the case of bremsstrahlung-like processes the
matrix element employed in BAMPS reads

\begin{equation}\label{gurnion-bertsch}
 \left | {\cal M}_{gg \to ggg} \right |^2 = \frac{72 \pi^2 \alpha^2_s s^2}{\left ( {\bf q}^2_\bot + m^2_D \right )^2} 
	  \frac{48 \pi \alpha_s {\bf q}^2_\bot  }
	  { {\bf k}^2_\bot \left [ \left ( {\bf k }_\bot - {\bf q}_\bot \right )^2 +
	  m^2_D \right ] } \ \Theta \left ( \Lambda_g - \tau \right )
\end{equation}
${\bf q}_\bot$ and ${\bf k }_\bot$ denote the perpendicular components of the
momentum transfer and of the radiated gluon momentum in the center of momentum
(CM) frame of the colliding particles, respectively.

When considering bremsstrahlung processes the LPM-effect \citep{LPM}, a
coherence effect named after Landau, Pomeranchuk and Migdal, needs to be taken
into account that leads to a suppression of the emission rate for low $p_\bot$
particles. Since such an interference effect cannot be incorporated directly
into a semi–classical microscopic transport model such as BAMPS, we choose an
effective approach by introducing the Theta function in (\ref{gurnion-bertsch}).
This implies that the formation time τ of the emitted gluon must not exceed the
mean free path of the parent gluon Λg , ensuring that successive radiative
processes are independent of each other.

We calculate the shear viscosity for a temperature of $T=400$ MeV.
Taking the coupling $\alpha_s$ as constant, the shear viscosity to
the entropy density ratio $\eta/s$ will not depend on temperature for the 
considered gluon interactions in our case \citep{PhysRevLett.100.172301}
(see later in Sec. \ref{angular_dependence}).
From next-to-leading-log calculation the shear viscosity
should roughly scale with $\eta_{NNL} \sim g^{-4} T^3 $ (though it has an
additional logarithmic dependence on $T$ due to Debye screening mass)
\citep{1126-6708-2003-05-051}
while the entropy of a quark gluon plasma should scale in leading order
with $s \sim T^3$ \citep{PhysRevLett.83.2906}, which implies a very weak
temperature dependence for $T=300MeV \ldots 600 MeV$

\begin{figure}[h]
  \includegraphics[width=0.9 \textwidth]{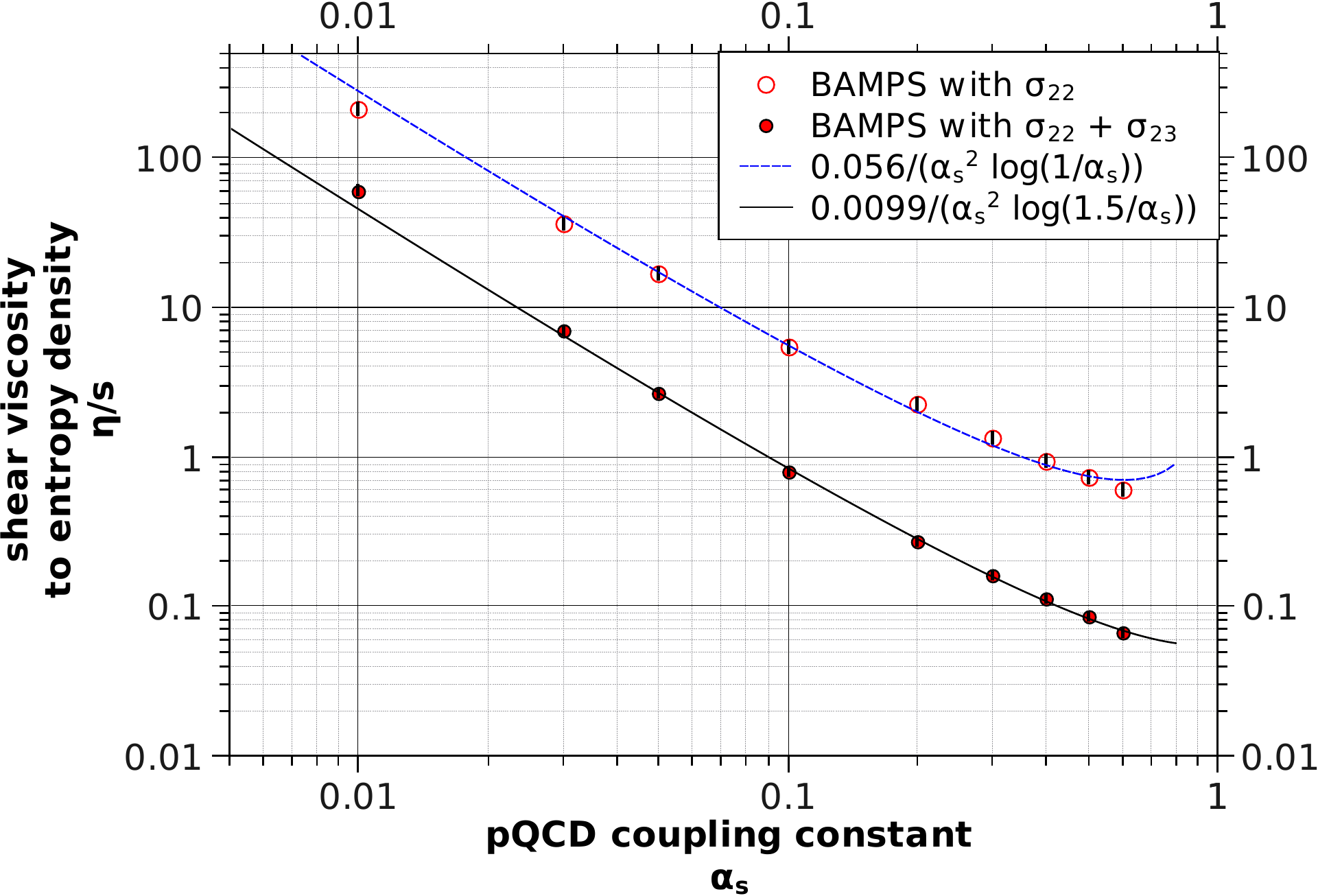} 
  \caption{(Color online) Shear viscosity to entropy density ratio for a gluon gas with
    pQCD-based interactions.}
  \label{fig:eta_s_pQCD}
\end{figure}

Figure \ref{fig:eta_s_pQCD} and Table \ref{table:alpha_s_1} present
the $\eta/s$ ratio of a gluon gas with the pQCD-based interactions, which
are the main results of this paper. Because gluons are considered as
Boltzmann particles, we have $s=4n=4d_GT^3/\pi^2$.

\begin{table}[h]
  \begin{tabular}{ccc}
    \hline \hline
    $\alpha_s$ 	& $\frac{\eta}{s}(gg \leftrightarrow gg)$\qquad	&
    $\frac{\eta}{s}(gg \leftrightarrow gg \ \& \ gg \leftrightarrow ggg)$
    \\
    \hline
    $0.01$	& $215 \pm 19$		&       $60 \pm 5$\\
    $0.03$	& $36.4 \pm 1.5$	&	$6.9 \pm 0.3$\\
    $0.05$	& $17.04 \pm 0.73$	&	$2.7 \pm 0.1$\\
    $0.1$	& $5.51 \pm 0.17$ 	&	$0.795 \pm 0.025$\\
    $0.2$	& $2.21 \pm 0.07$	&	$0.28 \pm 0.04$\\
    $0.3$	& $1.342 \pm 0.035$	&	$0.166 \pm 0.025$\\
    $0.4$	& $0.952 \pm 0.03$	&	$0.114 \pm 0.004$\\
    $0.5$	& $0.74 \pm 0.02$	&	$0.087 \pm 0.004$\\
    $0.6$	& $0.61 \pm 0.02$	&	$0.0664 \pm 0.002$\\
    \hline
  \end{tabular}
  \caption{Data plotted in Fig. \ref{fig:eta_s_pQCD}.}
  \label{table:alpha_s_1}
\end{table}

For pure gluon elastic interactions the fit function
(dashed curve) agrees with the analytical result 
of \cite{PhysRevLett.64.1867,Arnold:2000dr} at small $\alpha_s$.
Other numerical results from 
\citep{Bluhm:2010qf}
confirm this $\left ( \alpha_s^2\log(1/\alpha_s)\right )^{-1}$ scaling.
However, in \citep{Peshier:2005pp,Khvorostukhin:2010cw,Plumari:2011mk} a
different scaling $\eta \sim \left (\alpha_s\log(1/\alpha_s)\right )^{-1}$ was suggested,
which is a clear difference to the above findings. This different behavior stems
from the ansatz that the collision width $\tau^{-1} \sim \alpha_s T
\log(1/\alpha_s)$ is employed in their treatment of a simplified Boltzmann
collision process in the relaxation time approximation.

If the gluon bremsstrahlung $gg \leftrightarrow ggg$ is included, the shear
viscosity decreases by a factor of 7-8 for $\alpha_s > 0.1$ and a factor of
$3-6$ for $\alpha_s < 0.1$. The ratio of $\eta_{22}/\eta_{22+23}$ is much larger than
that of \citep{1126-6708-2003-05-051}. The reason was already pointed out in
\citep{PhysRevLett.100.172301}. While in the $\alpha_s \to 0$ limit the
collinear elastic processes are clearly dominant, for non-vanishing $\alpha_s$
the gluon radiation prefers rather large angles in the Bethe-Heitler regime
\citep{Fochler:2010wn}. In gluon radiation processes the distribution of the
radiation angle is close to isotropic \citep{PhysRevC.76.024911,Fochler:2010wn}.
For the $\eta/s$ ratio implementing both elastic and inelastic processes the fit
function is found to be $0.0099 \left (\alpha_s^2 \ \log(1.5/\alpha_s) \right
)^{-1}$ and is shown by the solid curve in Fig. \ref{fig:eta_s_pQCD}. This
scaling holds for rather small values of $\alpha_s$. In the limit of
$\alpha_s \to 0$ where bremsstrahlung becomes nearly collinear a scaling $\left
(\alpha_s^3 \ \log(1/\alpha_s) \right )^{-1}$ is expected for the inelastic
contributions \citep{PhysRevLett.100.172301}. 

\begin{figure}
  \includegraphics[width=0.9 \textwidth]{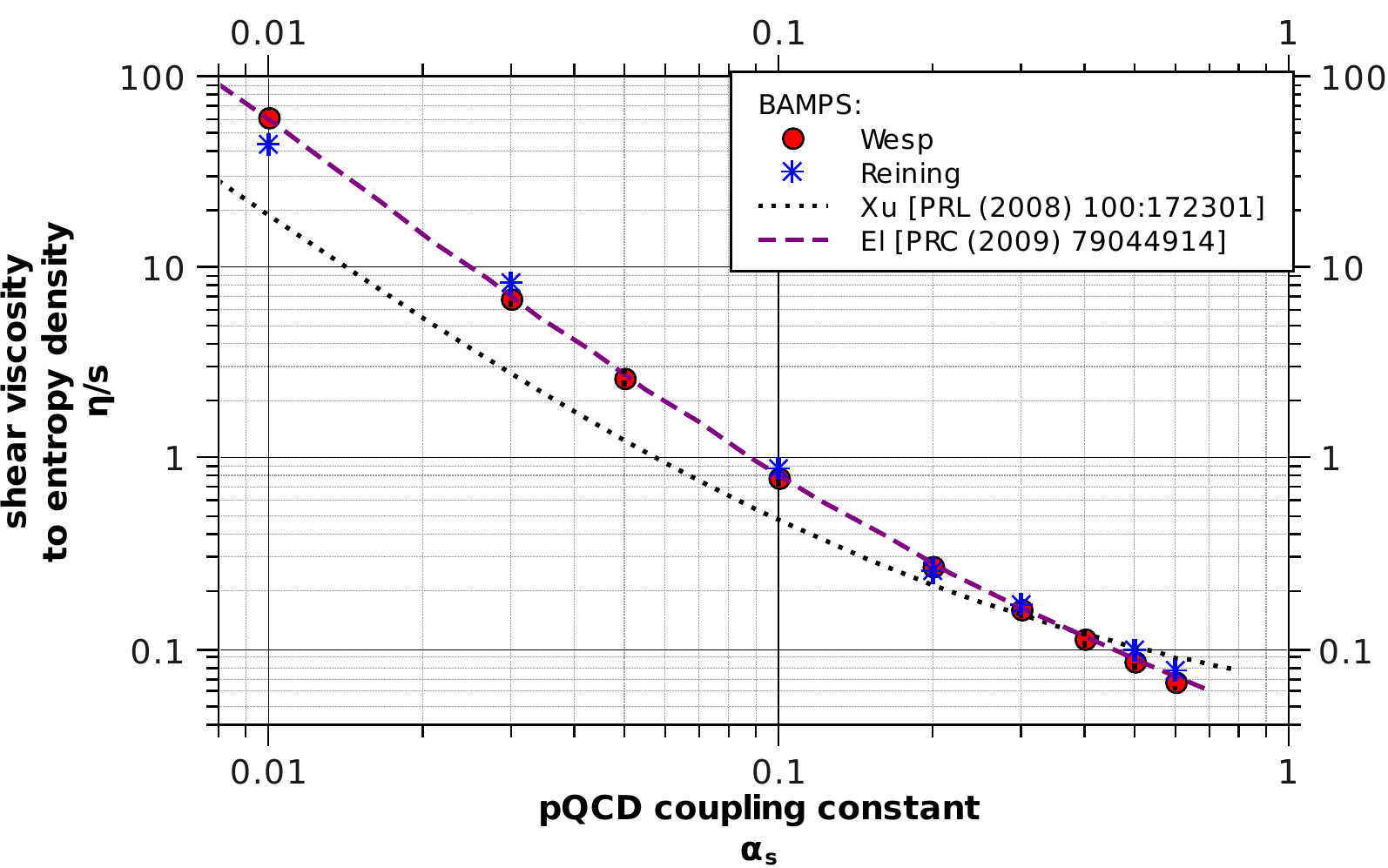} 
  \caption{(Color online) Comparisons of the results in Fig. \ref{fig:eta_s_pQCD} with
those of Refs. \citep{PhysRevLett.100.172301,El:2008yy,reining_paper}.}
\label{fig:eta_s_wg_vergleich}
\end{figure}

Figure \ref{fig:eta_s_wg_vergleich} shows comparisons of the present results
with those previously reported by some of the authors. The dotted curve is taken
from \citep{PhysRevLett.100.172301} and is obtained by using a gradient
expansion in the linearized Boltzmann equation. The deviation from thermal
equilibrium $\phi$, which is needed to calculate the shear viscosity, is
approximated by $\phi \sim (\chi/T)( p_z^2/E)$, where $\chi$ is a constant.
Actually, $\chi$ could be a function of momentum as indicated in the Grad's
ansatz, $\phi \sim C_0 \pi_{\mu\nu}p^\mu p^\nu$, where $C_0$ is a function of
temperature. It seems that the simple choice of $\phi$ in
\citep{PhysRevLett.100.172301} makes the $\eta/s$ values lower than the present
results at small $\alpha_s$. From Fig. \ref{fig:eta_s_wg_vergleich} we see that
the present results have a perfect agreement with those from \citep{El:2008yy}
(dashed curve). In \citep{El:2008yy} the shear viscosity is derived by
identifying the entropy production in hydrodynamic (Israel-Stewart) and kinetic
(Boltzmann) approach. For the deviation from thermal equilibrium the Grad's
ansatz is used in \citep{El:2008yy}.
In Fig. \ref{fig:eta_s_wg_vergleich} we also show new results (star symbols)
from another work done by Reining et al. \citep{reining_paper}.
There the shear viscosity is calculated as a ratio of the shear pressure
to the velocity gradient in a stationary shear flow pattern.
These new results agree well with the present ones. In summary, as noted earlier
in \citep{PhysRevLett.100.172301}, pQCD inelastic interactions can
drive the gluon matter to a strongly coupled system with an $\eta/s$ ratio as
small as the lower bound from the AdS-CFT conjecture.

\subsection{Temperature and collision angle dependence of the shear viscosity}
\label{angular_dependence}

\begin{figure}[h]
  \includegraphics[width=0.7 \textwidth]{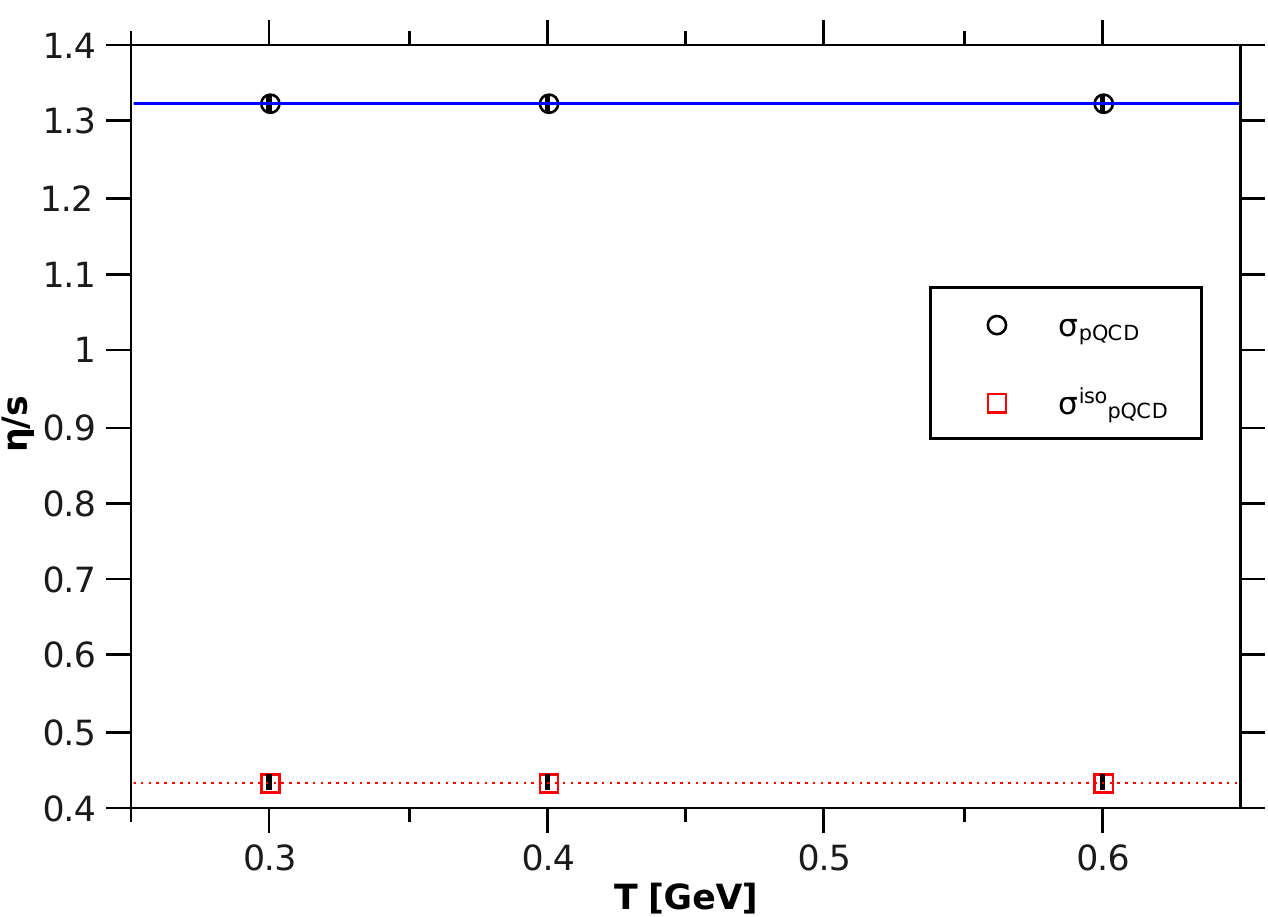}
  \caption{(Color online) Temperature dependence of the $\eta/s$ ratio. PQCD (circles)
and isotropic (squares) distribution of the collision angle in elastic
gluon interactions are taken for comparisons.}
  \label{fig:cs_comparisons}
\end{figure}

In this subsection we study the dependence of the shear viscosity on
temperature and the distribution of the collision angle. We only consider
elastic gluon collisions in order to repeat the calculations
performed in \citep{Fuini:2010xz}. Our results are shown in 
Fig. \ref{fig:cs_comparisons} and show some discrepancies with those of
\citep{Fuini:2010xz} at lower temperatures. At temperatures above $400$ MeV our
results are rather close to theirs.

As already done in the calculations presented in the previous
subsection, the cross section of
elastic gluon interactions is taken in the small-angle approximation
(\ref{csgg}). The $\eta/s$ ratios for the elastic gluon interactions
at $\alpha_s=0.3$ and at three chosen temperatures are shown in 
Fig. \ref{fig:cs_comparisons} by the circles. They agree well with 
a constant indicated by the line.
There is no temperature dependence of the shear viscosity to entropy ration
$\eta/s$ for the considered interactions, which is indicated in the findings
of \citep{Fuini:2010xz}. The authors of \citep{Fuini:2010xz} employed the same
cross section Eq. (\ref{csgg}) and calculated the $\eta/s$ ratio according to
the Green-Kubo relation (\ref{green_kubo_definition}) by using 
a transport model, the BMS implementation of PCM \citep{Bass:2002fh}.
They found an increase of $\eta/s$ for decreasing temperature.

From Eq. (\ref{Green_Kubo_simple_1}) the shear viscosity $\eta$ is proportional
to the product of the energy density $e$ and the relaxation time $\tau$.
It is clear that the relaxation time should be a product of the
mean free path $\lambda_{mfp}$ and a function of the distribution
of the collision angle $F[d(\theta)]$, i.e.,
$\tau \sim \lambda_{mfp} F[d(\theta)]$. From Eq. (\ref{csgg}) we obtain
\begin{eqnarray}
\label{csggtotal}
&& \sigma_{gg\to gg} =  \frac{9\pi\alpha_s^2}{2}
\frac{1}{m_D^2 \left ( 1+ 4 m_D^2/{\hat s} \right )} \\
\label{csggangle}
&& d(\theta)=\frac{1}{\sigma_{gg\to gg}} \frac{d\sigma_{gg\to gg}}{d\theta}=
\frac{4 \sin(2\theta) m_D^2 \left ( {\hat s} + 4m_D^2 \right )}
{\left ( {\hat s} \sin^2\theta+4m_D^2 \right )^2} \,.
\end{eqnarray}
Because momenta of gluons are distributed with $f(p) \sim e^{-E/T}$,
it is obvious that ${\hat s}$ scales with $T^2$, 
$\sigma_{gg\to gg}$ with $1/T^2$, and $d(\theta)$ has no dependence on $T$.
The mean free path $\lambda_{mfp}=1/n\langle \sigma_{gg\to gg} \rangle$
is then proportional to $1/T$, and $\eta \sim e \lambda_{mfp} F[d(\theta)]$
is proportional to $T^3$, like $s$. Thus, $\eta/s$ does not depend on
the temperature for the case Eq. (\ref{csgg}).

We now turn to discuss the dependence of the shear viscosity on the 
distribution of the collision angle. In analogy to the
procedure in \citep{Fuini:2010xz} we use the total cross section
(\ref{csggtotal}) to determine the collision probability
(\ref{eq_BAMPS_collProb}) and then replace the angular distribution
(\ref{csggangle}) by an isotropic form $d_{iso}(\theta) \sim \sin\theta$. 
The results on $\eta/s$ are also shown in Fig. \ref{fig:cs_comparisons}. The
$\eta/s$ ratio is smaller by a factor of $3$ if the pQCD angular distribution is replaced
by an isotropic one. 

To understand the dependence of the shear viscosity on the angular distribution,
we calculate the transport cross section, $\sigma^{tr}=\int d\theta \sin^2\theta
d\sigma/d\theta$, which is roughly inversely proportional to $\eta$. For the
isotropic angular distribution we have $\sigma^{tr}=(2/3) \sigma$, while for the
pQCD angular distribution
\begin{equation}
\frac{\sigma_{gg\to gg}^{tr}}{\sigma_{gg\to gg}}=
a(1+a)\left [ \ln(1+1/a) -1/(1+a) \right ]
\end{equation}
with $a=4m_D^2/{\hat s}$. On the average 
$\langle {\hat s} \rangle=18 T^2$ and thus 
$\langle a \rangle=16 \alpha_s/(3\pi)\approx 0.5$ for $\alpha_s=0.3$.
We obtain $\sigma_{gg\to gg}^{tr}=0.32 \sigma_{gg\to gg}$, which is
a factor of $2$ smaller than that using the isotropic angular distribution.
By this analytical estimation the shear viscosity effectively decreases by
a factor of two if one replaces the forward-backward
peaked pQCD angular distribution by an isotropic one. From numerical
calculations we see a factor of $3$ as seen in Fig. \ref{fig:cs_comparisons}.
This situation for elastic scatterings also indicates the efficiency of gluon
bremsstrahlung $gg\leftrightarrow ggg$ in thermalization and lowering the shear
viscosity if the fact of the finite radiation angle at moderate $\alpha_s$ is
taken into account.

\section{Summary}
In this work we calculated the shear viscosity with the Green-Kubo relation. The
correlation function is extracted from simulations by the parton cascade BAMPS.
To cross-check the numerical method, the shear viscosity for interactions with
constant cross section and isotropic angular distribution is calculated and the
results agree perfectly with the analytical ones. The main results are the
values of $\eta/s$ ratio of a gluon gas including a $2 \leftrightarrow 3$
pQCD-based gluon bremsstrahlung process. These results are in perfect agreement
with previously published numbers employing other methods. In comparison to pure
elastic processes the the $\eta/s$ ratio is lowered by a factor of $3 - 8$ for
$\alpha_s=0.01 - 0.6$ compared to solely elastic collisions when gluon
bremsstrahlung is included. A nice scaling of $\eta \sim 1/[\alpha_s^2 \ln
(1.5/\alpha_s)]$ in the coupling constant $\alpha_s$ is found as expected from
standard pQCD arguments. We also showed that $\eta/s$ has no temperature
dependence at constant $\alpha_s$ and the change of the pQCD elastic
interactions to those with isotropic angular distribution brings a factor of $3$
decrease in the shear viscosity.

\section*{Acknowledgements} The authors are grateful to the Center for the
Scientific Computing (CSC) at Frankfurt for the computing resources. C. Wesp,
C.Greiner and Z. Xu thank S. Bass for fruitful discussions. C. Wesp, F. Reining,
A. El and I. Bouras are grateful to HGS-HIRe. C. Wesp and A. El are grateful to
H-QM. This work was supported by the Helmholtz International Center for FAIR
within the framework of the LOEWE program launched by the State of Hessen.

\nocite{*}

\bibliography{paper}

\begin{thebibliography}{54}%
\makeatletter
\providecommand \@ifxundefined [1]{%
 \@ifx{#1\undefined}
}%
\providecommand \@ifnum [1]{%
 \ifnum #1\expandafter \@firstoftwo
 \else \expandafter \@secondoftwo
 \fi
}%
\providecommand \@ifx [1]{%
 \ifx #1\expandafter \@firstoftwo
 \else \expandafter \@secondoftwo
 \fi
}%
\providecommand \natexlab [1]{#1}%
\providecommand \enquote  [1]{``#1''}%
\providecommand \bibnamefont  [1]{#1}%
\providecommand \bibfnamefont [1]{#1}%
\providecommand \citenamefont [1]{#1}%
\providecommand \href@noop [0]{\@secondoftwo}%
\providecommand \href [0]{\begingroup \@sanitize@url \@href}%
\providecommand \@href[1]{\@@startlink{#1}\@@href}%
\providecommand \@@href[1]{\endgroup#1\@@endlink}%
\providecommand \@sanitize@url [0]{\catcode `\\12\catcode `\$12\catcode
  `\&12\catcode `\#12\catcode `\^12\catcode `\_12\catcode `\%12\relax}%
\providecommand \@@startlink[1]{}%
\providecommand \@@endlink[0]{}%
\providecommand \url  [0]{\begingroup\@sanitize@url \@url }%
\providecommand \@url [1]{\endgroup\@href {#1}{\urlprefix }}%
\providecommand \urlprefix  [0]{URL }%
\providecommand \Eprint [0]{\href }%
\providecommand \doibase [0]{http://dx.doi.org/}%
\providecommand \selectlanguage [0]{\@gobble}%
\providecommand \bibinfo  [0]{\@secondoftwo}%
\providecommand \bibfield  [0]{\@secondoftwo}%
\providecommand \translation [1]{[#1]}%
\providecommand \BibitemOpen [0]{}%
\providecommand \bibitemStop [0]{}%
\providecommand \bibitemNoStop [0]{.\EOS\space}%
\providecommand \EOS [0]{\spacefactor3000\relax}%
\providecommand \BibitemShut  [1]{\csname bibitem#1\endcsname}%
\let\auto@bib@innerbib\@empty
\bibitem [{\citenamefont {Ackermann}\ \emph {et~al.}(2001)\citenamefont
  {Ackermann} \emph {et~al.}}]{Ackermann:2000tr}%
  \BibitemOpen
  \bibfield  {author} {\bibinfo {author} {\bibfnamefont {K.~H.}\ \bibnamefont
  {Ackermann}} \emph {et~al.},\ }\href@noop {} {\bibfield  {journal} {\bibinfo
  {journal} {Phys. Rev. Lett.}\ }\textbf {\bibinfo {volume} {86}},\ \bibinfo
  {pages} {402} (\bibinfo {year} {2001})}\BibitemShut {NoStop}%
\bibitem [{\citenamefont {Adler}\ \emph {et~al.}(2003)\citenamefont {Adler}
  \emph {et~al.}}]{PhysRevLett.91.182301}%
  \BibitemOpen
  \bibfield  {author} {\bibinfo {author} {\bibfnamefont {S.~S.}\ \bibnamefont
  {Adler}} \emph {et~al.} (\bibinfo {collaboration} {PHENIX Collaboration}),\
  }\href@noop {} {\bibfield  {journal} {\bibinfo  {journal} {Phys. Rev. Lett.}\
  }\textbf {\bibinfo {volume} {91}},\ \bibinfo {pages} {182301} (\bibinfo
  {year} {2003})}\BibitemShut {NoStop}%
\bibitem [{\citenamefont {Adams}\ \emph {et~al.}(2004)\citenamefont {Adams}
  \emph {et~al.}}]{Adams:2003am}%
  \BibitemOpen
  \bibfield  {author} {\bibinfo {author} {\bibfnamefont {J.}~\bibnamefont
  {Adams}} \emph {et~al.} (\bibinfo {collaboration} {STAR Collaboration}),\
  }\href@noop {} {\bibfield  {journal} {\bibinfo  {journal} {Phys. Rev. Lett.}\
  }\textbf {\bibinfo {volume} {92}},\ \bibinfo {pages} {052302} (\bibinfo
  {year} {2004})}\BibitemShut {NoStop}%
\bibitem [{\citenamefont {Back}\ \emph {et~al.}(2005)\citenamefont {Back} \emph
  {et~al.}}]{PhysRevLett.94.122303}%
  \BibitemOpen
  \bibfield  {author} {\bibinfo {author} {\bibfnamefont {B.~B.}\ \bibnamefont
  {Back}} \emph {et~al.},\ }\href@noop {} {\bibfield  {journal} {\bibinfo
  {journal} {Phys. Rev. Lett.}\ }\textbf {\bibinfo {volume} {94}},\ \bibinfo
  {pages} {122303} (\bibinfo {year} {2005})}\BibitemShut {NoStop}%
\bibitem [{\citenamefont {Alver}\ \emph {et~al.}(2007)\citenamefont {Alver}
  \emph {et~al.}}]{PhysRevLett.98.242302}%
  \BibitemOpen
  \bibfield  {author} {\bibinfo {author} {\bibfnamefont {B.}~\bibnamefont
  {Alver}} \emph {et~al.},\ }\href@noop {} {\bibfield  {journal} {\bibinfo
  {journal} {Phys. Rev. Lett.}\ }\textbf {\bibinfo {volume} {98}},\ \bibinfo
  {pages} {242302} (\bibinfo {year} {2007})}\BibitemShut {NoStop}%
\bibitem [{\citenamefont {Afanasiev}\ \emph {et~al.}(2009)\citenamefont
  {Afanasiev} \emph {et~al.}}]{PhysRevC.80.024909}%
  \BibitemOpen
  \bibfield  {author} {\bibinfo {author} {\bibfnamefont {S.}~\bibnamefont
  {Afanasiev}} \emph {et~al.} (\bibinfo {collaboration} {PHENIX
  Collaboration}),\ }\href@noop {} {\bibfield  {journal} {\bibinfo  {journal}
  {Phys. Rev. C}\ }\textbf {\bibinfo {volume} {80}},\ \bibinfo {pages} {024909}
  (\bibinfo {year} {2009})}\BibitemShut {NoStop}%
\bibitem [{\citenamefont {Aamodt}\ \emph {et~al.}(2010)\citenamefont {Aamodt}
  \emph {et~al.}}]{lhc}%
  \BibitemOpen
  \bibfield  {author} {\bibinfo {author} {\bibfnamefont {K.}~\bibnamefont
  {Aamodt}} \emph {et~al.} (\bibinfo {collaboration} {ALICE Collaboration}),\
  }\href@noop {} {\  (\bibinfo {year} {2010})},\ \Eprint
  {http://arxiv.org/abs/1011.3914} {arXiv:1011.3914 [nucl-ex]} \BibitemShut
  {NoStop}%
\bibitem [{\citenamefont {Song}\ and\ \citenamefont
  {Heinz}(2008)}]{Song:2007ux}%
  \BibitemOpen
  \bibfield  {author} {\bibinfo {author} {\bibfnamefont {H.}~\bibnamefont
  {Song}}\ and\ \bibinfo {author} {\bibfnamefont {U.~W.}\ \bibnamefont
  {Heinz}},\ }\href@noop {} {\bibfield  {journal} {\bibinfo  {journal}
  {Phys.Rev.}\ }\textbf {\bibinfo {volume} {C77}},\ \bibinfo {pages} {064901}
  (\bibinfo {year} {2008})}\BibitemShut {NoStop}%
\bibitem [{\citenamefont {Luzum}\ and\ \citenamefont
  {Romatschke}(2008)}]{PhysRevC.78.034915}%
  \BibitemOpen
  \bibfield  {author} {\bibinfo {author} {\bibfnamefont {M.}~\bibnamefont
  {Luzum}}\ and\ \bibinfo {author} {\bibfnamefont {P.}~\bibnamefont
  {Romatschke}},\ }\href@noop {} {\bibfield  {journal} {\bibinfo  {journal}
  {Phys. Rev. C}\ }\textbf {\bibinfo {volume} {78}},\ \bibinfo {pages} {034915}
  (\bibinfo {year} {2008})}\BibitemShut {NoStop}%
\bibitem [{\citenamefont {Teaney}(2009)}]{Teaney:2009qa}%
  \BibitemOpen
  \bibfield  {author} {\bibinfo {author} {\bibfnamefont {D.~A.}\ \bibnamefont
  {Teaney}},\ }\href@noop {} {\  (\bibinfo {year} {2009})},\ \Eprint
  {http://arxiv.org/abs/0905.2433} {arXiv:0905.2433 [nucl-th]} \BibitemShut
  {NoStop}%
\bibitem [{\citenamefont {Schenke}\ \emph {et~al.}(2010)\citenamefont
  {Schenke}, \citenamefont {Jeon},\ and\ \citenamefont
  {Gale}}]{Schenke:2010nt}%
  \BibitemOpen
  \bibfield  {author} {\bibinfo {author} {\bibfnamefont {B.}~\bibnamefont
  {Schenke}}, \bibinfo {author} {\bibfnamefont {S.}~\bibnamefont {Jeon}}, \
  and\ \bibinfo {author} {\bibfnamefont {C.}~\bibnamefont {Gale}},\ }\href@noop
  {} {\bibfield  {journal} {\bibinfo  {journal} {Phys.Rev.}\ }\textbf {\bibinfo
  {volume} {C82}},\ \bibinfo {pages} {014903} (\bibinfo {year}
  {2010})}\BibitemShut {NoStop}%
\bibitem [{\citenamefont {Song}\ \emph {et~al.}(2011)\citenamefont {Song},
  \citenamefont {Bass},\ and\ \citenamefont {Heinz}}]{Song:2011qa}%
  \BibitemOpen
  \bibfield  {author} {\bibinfo {author} {\bibfnamefont {H.}~\bibnamefont
  {Song}}, \bibinfo {author} {\bibfnamefont {S.~A.}\ \bibnamefont {Bass}}, \
  and\ \bibinfo {author} {\bibfnamefont {U.~W.}\ \bibnamefont {Heinz}},\
  }\href@noop {} {\  (\bibinfo {year} {2011})},\ \Eprint
  {http://arxiv.org/abs/1103.2380} {arXiv:1103.2380 [nucl-th]} \BibitemShut
  {NoStop}%
\bibitem [{\citenamefont {Schenke}\ \emph {et~al.}(2011)\citenamefont
  {Schenke}, \citenamefont {Jeon},\ and\ \citenamefont
  {Gale}}]{Schenke:2011tv}%
  \BibitemOpen
  \bibfield  {author} {\bibinfo {author} {\bibfnamefont {B.}~\bibnamefont
  {Schenke}}, \bibinfo {author} {\bibfnamefont {S.}~\bibnamefont {Jeon}}, \
  and\ \bibinfo {author} {\bibfnamefont {C.}~\bibnamefont {Gale}},\ }\href@noop
  {} {\  (\bibinfo {year} {2011})},\ \Eprint {http://arxiv.org/abs/1102.0575}
  {arXiv:1102.0575 [hep-ph]} \BibitemShut {NoStop}%
\bibitem [{\citenamefont {Xu}\ and\ \citenamefont
  {Greiner}(2005)}]{PhysRevC.71.064901}%
  \BibitemOpen
  \bibfield  {author} {\bibinfo {author} {\bibfnamefont {Z.}~\bibnamefont
  {Xu}}\ and\ \bibinfo {author} {\bibfnamefont {C.}~\bibnamefont {Greiner}},\
  }\href@noop {} {\bibfield  {journal} {\bibinfo  {journal} {Phys. Rev. C}\
  }\textbf {\bibinfo {volume} {71}},\ \bibinfo {pages} {064901} (\bibinfo
  {year} {2005})}\BibitemShut {NoStop}%
\bibitem [{\citenamefont {Xu.}\ \emph {et~al.}(2008)\citenamefont {Xu.},
  \citenamefont {Greiner},\ and\ \citenamefont {St\"ocker}}]{Xu2008a}%
  \BibitemOpen
  \bibfield  {author} {\bibinfo {author} {\bibfnamefont {Z.}~\bibnamefont
  {Xu.}}, \bibinfo {author} {\bibfnamefont {C.}~\bibnamefont {Greiner}}, \ and\
  \bibinfo {author} {\bibfnamefont {H.}~\bibnamefont {St\"ocker}},\ }\href@noop
  {} {\bibfield  {journal} {\bibinfo  {journal} {Phys. Rev. Lett.}\ }\textbf
  {\bibinfo {volume} {101}},\ \bibinfo {pages} {082302} (\bibinfo {year}
  {2008})}\BibitemShut {NoStop}%
\bibitem [{\citenamefont {Ferini}\ \emph {et~al.}(2009)\citenamefont {Ferini},
  \citenamefont {Colonna}, \citenamefont {Di~Toro},\ and\ \citenamefont
  {Greco}}]{Ferini:2008he}%
  \BibitemOpen
  \bibfield  {author} {\bibinfo {author} {\bibfnamefont {G.}~\bibnamefont
  {Ferini}}, \bibinfo {author} {\bibfnamefont {M.}~\bibnamefont {Colonna}},
  \bibinfo {author} {\bibfnamefont {M.}~\bibnamefont {Di~Toro}}, \ and\
  \bibinfo {author} {\bibfnamefont {V.}~\bibnamefont {Greco}},\ }\href@noop {}
  {\bibfield  {journal} {\bibinfo  {journal} {Phys. Lett.}\ }\textbf {\bibinfo
  {volume} {B670}},\ \bibinfo {pages} {325} (\bibinfo {year}
  {2009})}\BibitemShut {NoStop}%
\bibitem [{\citenamefont {Molnar}\ and\ \citenamefont
  {Gyulassy}(2002)}]{Molnar:2001nk}%
  \BibitemOpen
  \bibfield  {author} {\bibinfo {author} {\bibfnamefont {D.}~\bibnamefont
  {Molnar}}\ and\ \bibinfo {author} {\bibfnamefont {M.}~\bibnamefont
  {Gyulassy}},\ }\href@noop {} {\bibfield  {journal} {\bibinfo  {journal}
  {Nucl. Phys.}\ }\textbf {\bibinfo {volume} {A698}},\ \bibinfo {pages} {379}
  (\bibinfo {year} {2002})}\BibitemShut {NoStop}%
\bibitem [{\citenamefont {Arnold}\ \emph {et~al.}(2003)\citenamefont {Arnold},
  \citenamefont {Moore},\ and\ \citenamefont {Yaffe}}]{1126-6708-2003-05-051}%
  \BibitemOpen
  \bibfield  {author} {\bibinfo {author} {\bibfnamefont {P.}~\bibnamefont
  {Arnold}}, \bibinfo {author} {\bibfnamefont {G.~D.}\ \bibnamefont {Moore}}, \
  and\ \bibinfo {author} {\bibfnamefont {L.~G.}\ \bibnamefont {Yaffe}},\
  }\href@noop {} {\bibfield  {journal} {\bibinfo  {journal} {Journal of High
  Energy Physics}\ }\textbf {\bibinfo {volume} {2003}},\ \bibinfo {pages} {051}
  (\bibinfo {year} {2003})}\BibitemShut {NoStop}%
\bibitem [{\citenamefont {Kovtun}\ \emph {et~al.}(2005)\citenamefont {Kovtun},
  \citenamefont {Son},\ and\ \citenamefont {Starinets}}]{Kovtun:2004de}%
  \BibitemOpen
  \bibfield  {author} {\bibinfo {author} {\bibfnamefont {P.}~\bibnamefont
  {Kovtun}}, \bibinfo {author} {\bibfnamefont {D.}~\bibnamefont {Son}}, \ and\
  \bibinfo {author} {\bibfnamefont {A.}~\bibnamefont {Starinets}},\ }\href@noop
  {} {\bibfield  {journal} {\bibinfo  {journal} {Phys.Rev.Lett.}\ }\textbf
  {\bibinfo {volume} {94}},\ \bibinfo {pages} {111601} (\bibinfo {year}
  {2005})}\BibitemShut {NoStop}%
\bibitem [{\citenamefont {Bluhm}\ \emph {et~al.}(2010)\citenamefont {Bluhm},
  \citenamefont {Kampfer},\ and\ \citenamefont {Redlich}}]{Bluhm:2010qf}%
  \BibitemOpen
  \bibfield  {author} {\bibinfo {author} {\bibfnamefont {M.}~\bibnamefont
  {Bluhm}}, \bibinfo {author} {\bibfnamefont {B.}~\bibnamefont {Kampfer}}, \
  and\ \bibinfo {author} {\bibfnamefont {K.}~\bibnamefont {Redlich}},\
  }\href@noop {} {\bibfield  {journal} {\bibinfo  {journal} {arXiv:1011.5634
  [hep-ph]}\ } (\bibinfo {year} {2010})}\BibitemShut {NoStop}%
\bibitem [{\citenamefont {Xu}\ and\ \citenamefont
  {Greiner}(2008)}]{PhysRevLett.100.172301}%
  \BibitemOpen
  \bibfield  {author} {\bibinfo {author} {\bibfnamefont {Z.}~\bibnamefont
  {Xu}}\ and\ \bibinfo {author} {\bibfnamefont {C.}~\bibnamefont {Greiner}},\
  }\href@noop {} {\bibfield  {journal} {\bibinfo  {journal} {Phys. Rev. Lett.}\
  }\textbf {\bibinfo {volume} {100}},\ \bibinfo {pages} {172301} (\bibinfo
  {year} {2008})}\BibitemShut {NoStop}%
\bibitem [{\citenamefont {El}\ \emph {et~al.}(2009)\citenamefont {El},
  \citenamefont {Muronga}, \citenamefont {Xu},\ and\ \citenamefont
  {Greiner}}]{El:2008yy}%
  \BibitemOpen
  \bibfield  {author} {\bibinfo {author} {\bibfnamefont {A.}~\bibnamefont
  {El}}, \bibinfo {author} {\bibfnamefont {A.}~\bibnamefont {Muronga}},
  \bibinfo {author} {\bibfnamefont {Z.}~\bibnamefont {Xu}}, \ and\ \bibinfo
  {author} {\bibfnamefont {C.}~\bibnamefont {Greiner}},\ }\href@noop {}
  {\bibfield  {journal} {\bibinfo  {journal} {Phys. Rev.}\ }\textbf {\bibinfo
  {volume} {C79}},\ \bibinfo {pages} {044914} (\bibinfo {year}
  {2009})}\BibitemShut {NoStop}%
\bibitem [{\citenamefont {Chen}\ \emph {et~al.}(2010)\citenamefont {Chen},
  \citenamefont {Dong}, \citenamefont {Ohnishi},\ and\ \citenamefont
  {Wang}}]{Chen:2009sm}%
  \BibitemOpen
  \bibfield  {author} {\bibinfo {author} {\bibfnamefont {J.-W.}\ \bibnamefont
  {Chen}}, \bibinfo {author} {\bibfnamefont {H.}~\bibnamefont {Dong}}, \bibinfo
  {author} {\bibfnamefont {K.}~\bibnamefont {Ohnishi}}, \ and\ \bibinfo
  {author} {\bibfnamefont {Q.}~\bibnamefont {Wang}},\ }\href@noop {} {\bibfield
   {journal} {\bibinfo  {journal} {Phys. Lett.}\ }\textbf {\bibinfo {volume}
  {B685}},\ \bibinfo {pages} {277} (\bibinfo {year} {2010})}\BibitemShut
  {NoStop}%
\bibitem [{\citenamefont {Chen}\ \emph {et~al.}(2011)\citenamefont {Chen},
  \citenamefont {Deng}, \citenamefont {Dong},\ and\ \citenamefont
  {Wang}}]{Chen:2010xk}%
  \BibitemOpen
  \bibfield  {author} {\bibinfo {author} {\bibfnamefont {J.-W.}\ \bibnamefont
  {Chen}}, \bibinfo {author} {\bibfnamefont {J.}~\bibnamefont {Deng}}, \bibinfo
  {author} {\bibfnamefont {H.}~\bibnamefont {Dong}}, \ and\ \bibinfo {author}
  {\bibfnamefont {Q.}~\bibnamefont {Wang}},\ }\href@noop {} {\bibfield
  {journal} {\bibinfo  {journal} {Phys. Rev.}\ }\textbf {\bibinfo {volume}
  {D83}},\ \bibinfo {pages} {034031} (\bibinfo {year} {2011})}\BibitemShut
  {NoStop}%
\bibitem [{\citenamefont {Fuini}\ \emph {et~al.}(2011)\citenamefont {Fuini},
  \citenamefont {Demir}, \citenamefont {Srivastava},\ and\ \citenamefont
  {Bass}}]{Fuini:2010xz}%
  \BibitemOpen
  \bibfield  {author} {\bibinfo {author} {\bibfnamefont {J.}~\bibnamefont
  {Fuini}, \bibfnamefont {III}}, \bibinfo {author} {\bibfnamefont {N.~S.}\
  \bibnamefont {Demir}}, \bibinfo {author} {\bibfnamefont {D.~K.}\ \bibnamefont
  {Srivastava}}, \ and\ \bibinfo {author} {\bibfnamefont {S.~A.}\ \bibnamefont
  {Bass}},\ }\href@noop {} {\bibfield  {journal} {\bibinfo  {journal} {J.
  Phys.}\ }\textbf {\bibinfo {volume} {G38}},\ \bibinfo {pages} {015004}
  (\bibinfo {year} {2011})}\BibitemShut {NoStop}%
\bibitem [{\citenamefont {Green}(1954)}]{kubo1957statistical2}%
  \BibitemOpen
  \bibfield  {author} {\bibinfo {author} {\bibfnamefont {M.~S.}\ \bibnamefont
  {Green}},\ }\href@noop {} {\bibfield  {journal} {\bibinfo  {journal} {Journal
  of Chemical Physics}\ }\textbf {\bibinfo {volume} {22}},\ \bibinfo {pages}
  {398} (\bibinfo {year} {1954})}\BibitemShut {NoStop}%
\bibitem [{\citenamefont {Kubo}(1957)}]{kubo1957statistical}%
  \BibitemOpen
  \bibfield  {author} {\bibinfo {author} {\bibfnamefont {R.}~\bibnamefont
  {Kubo}},\ }\href@noop {} {\bibfield  {journal} {\bibinfo  {journal} {Journal
  of the Physical Society of Japan}\ }\textbf {\bibinfo {volume} {12}},\
  \bibinfo {pages} {570} (\bibinfo {year} {1957})}\BibitemShut {NoStop}%
\bibitem [{\citenamefont {Onsager}(1931)}]{onsager}%
  \BibitemOpen
  \bibfield  {author} {\bibinfo {author} {\bibfnamefont {L.}~\bibnamefont
  {Onsager}},\ }\href@noop {} {\bibfield  {journal} {\bibinfo  {journal}
  {PhysRev.37.405-426n}\ }\textbf {\bibinfo {volume} {37}},\ \bibinfo {pages}
  {405} (\bibinfo {year} {1931})}\BibitemShut {NoStop}%
\bibitem [{\citenamefont {Zubarev}(1996{\natexlab{a}})}]{book2}%
  \BibitemOpen
  \bibfield  {author} {\bibinfo {author} {\bibfnamefont {R.}~\bibnamefont
  {Zubarev}, \bibfnamefont {Morozov}},\ }\href@noop {} {\emph {\bibinfo {title}
  {Statistical Mechanics of Nonequilibrium Processes Volume 2: Relaxation and
  Hydrodynamic Processes}}}\ (\bibinfo  {publisher} {Akademie Verlag GmbH},\
  \bibinfo {year} {1996})\BibitemShut {NoStop}%
\bibitem [{\citenamefont {Zubarev}(1996{\natexlab{b}})}]{book1}%
  \BibitemOpen
  \bibfield  {author} {\bibinfo {author} {\bibfnamefont {R.}~\bibnamefont
  {Zubarev}, \bibfnamefont {Morozov}},\ }\href@noop {} {\emph {\bibinfo {title}
  {Statistical Mechanics of Nonequilibrium Processes Volume 1: Basic Concepts,
  Kinetic Theory}}}\ (\bibinfo  {publisher} {Akademie Verlag GmbH},\ \bibinfo
  {year} {1996})\BibitemShut {NoStop}%
\bibitem [{\citenamefont {Calzetta}(2008)}]{esterban_book}%
  \BibitemOpen
  \bibfield  {author} {\bibinfo {author} {\bibfnamefont {E.~A.}\ \bibnamefont
  {Calzetta}},\ }\href@noop {} {\emph {\bibinfo {title} {Nonequilibrium Quantum
  Field Theory}}}\ (\bibinfo  {publisher} {Cambridge University Press},\
  \bibinfo {year} {2008})\BibitemShut {NoStop}%
\bibitem [{\citenamefont {Searles}\ and\ \citenamefont
  {Evans}(2000)}]{searles:9727}%
  \BibitemOpen
  \bibfield  {author} {\bibinfo {author} {\bibfnamefont {D.~J.}\ \bibnamefont
  {Searles}}\ and\ \bibinfo {author} {\bibfnamefont {D.~J.}\ \bibnamefont
  {Evans}},\ }\href {http://link.aip.org/link/?JCP/112/9727/1} {\bibfield
  {journal} {\bibinfo  {journal} {The Journal of Chemical Physics}\ }\textbf
  {\bibinfo {volume} {112}},\ \bibinfo {pages} {9727} (\bibinfo {year}
  {2000})}\BibitemShut {NoStop}%
\bibitem [{\citenamefont {Bass}\ \emph {et~al.}(1998)\citenamefont {Bass} \emph
  {et~al.}}]{Bass:1998ca}%
  \BibitemOpen
  \bibfield  {author} {\bibinfo {author} {\bibfnamefont {S.~A.}\ \bibnamefont
  {Bass}} \emph {et~al.},\ }\href@noop {} {\bibfield  {journal} {\bibinfo
  {journal} {Prog. Part. Nucl. Phys.}\ }\textbf {\bibinfo {volume} {41}},\
  \bibinfo {pages} {255} (\bibinfo {year} {1998})}\BibitemShut {NoStop}%
\bibitem [{\citenamefont {Bleicher}\ \emph {et~al.}(1999)\citenamefont
  {Bleicher} \emph {et~al.}}]{Bleicher:1999xi}%
  \BibitemOpen
  \bibfield  {author} {\bibinfo {author} {\bibfnamefont {M.}~\bibnamefont
  {Bleicher}} \emph {et~al.},\ }\href@noop {} {\bibfield  {journal} {\bibinfo
  {journal} {J. Phys.}\ }\textbf {\bibinfo {volume} {G25}},\ \bibinfo {pages}
  {1859} (\bibinfo {year} {1999})}\BibitemShut {NoStop}%
\bibitem [{\citenamefont {Bass}\ \emph {et~al.}(2003)\citenamefont {Bass},
  \citenamefont {Muller},\ and\ \citenamefont {Srivastava}}]{Bass:2002fh}%
  \BibitemOpen
  \bibfield  {author} {\bibinfo {author} {\bibfnamefont {S.~A.}\ \bibnamefont
  {Bass}}, \bibinfo {author} {\bibfnamefont {B.}~\bibnamefont {Muller}}, \ and\
  \bibinfo {author} {\bibfnamefont {D.~K.}\ \bibnamefont {Srivastava}},\
  }\href@noop {} {\bibfield  {journal} {\bibinfo  {journal} {Phys. Lett.}\
  }\textbf {\bibinfo {volume} {B551}},\ \bibinfo {pages} {277} (\bibinfo {year}
  {2003})}\BibitemShut {NoStop}%
\bibitem [{\citenamefont {Demir}\ and\ \citenamefont
  {Bass}(2009)}]{Demir:2008tr}%
  \BibitemOpen
  \bibfield  {author} {\bibinfo {author} {\bibfnamefont {N.}~\bibnamefont
  {Demir}}\ and\ \bibinfo {author} {\bibfnamefont {S.~A.}\ \bibnamefont
  {Bass}},\ }\href@noop {} {\bibfield  {journal} {\bibinfo  {journal} {Phys.
  Rev. Lett.}\ }\textbf {\bibinfo {volume} {102}},\ \bibinfo {pages} {172302}
  (\bibinfo {year} {2009})}\BibitemShut {NoStop}%
\bibitem [{\citenamefont {Xu}\ and\ \citenamefont
  {Greiner}(2007)}]{PhysRevC.76.024911}%
  \BibitemOpen
  \bibfield  {author} {\bibinfo {author} {\bibfnamefont {Z.}~\bibnamefont
  {Xu}}\ and\ \bibinfo {author} {\bibfnamefont {C.}~\bibnamefont {Greiner}},\
  }\href@noop {} {\bibfield  {journal} {\bibinfo  {journal} {Phys. Rev. C}\
  }\textbf {\bibinfo {volume} {76}},\ \bibinfo {pages} {024911} (\bibinfo
  {year} {2007})}\BibitemShut {NoStop}%
\bibitem [{\citenamefont {Wesp}(2010)}]{wesp-master}%
  \BibitemOpen
  \bibfield  {author} {\bibinfo {author} {\bibfnamefont {C.}~\bibnamefont
  {Wesp}},\ }\emph {\bibinfo {title} {Shear viscosity, Green-Kubo relations and
  transport simulations - unpublished}},\ \href@noop {} {Master's thesis},\
  \bibinfo  {school} {Goethe-Universit\"{a}t Frankfurt am Main} (\bibinfo
  {year} {2010})\BibitemShut {NoStop}%
\bibitem [{\citenamefont {Reichl}(1980)}]{reichl_book}%
  \BibitemOpen
  \bibfield  {author} {\bibinfo {author} {\bibfnamefont {L.~E.}\ \bibnamefont
  {Reichl}},\ }\href@noop {} {\emph {\bibinfo {title} {A Modern Course in
  Statistical Mechanics}}}\ (\bibinfo  {publisher} {Hodder \& Stoughton
  Educational},\ \bibinfo {year} {1980})\BibitemShut {NoStop}%
\bibitem [{\citenamefont {Hirano}\ and\ \citenamefont
  {Gyulassy}(2006)}]{Hirano:2005wx}%
  \BibitemOpen
  \bibfield  {author} {\bibinfo {author} {\bibfnamefont {T.}~\bibnamefont
  {Hirano}}\ and\ \bibinfo {author} {\bibfnamefont {M.}~\bibnamefont
  {Gyulassy}},\ }\href@noop {} {\bibfield  {journal} {\bibinfo  {journal}
  {Nucl.Phys.}\ }\textbf {\bibinfo {volume} {A769}},\ \bibinfo {pages} {71}
  (\bibinfo {year} {2006})}\BibitemShut {NoStop}%
\bibitem [{\citenamefont {de~Groot}\ and\ \citenamefont {van
  Leeuwen}(1980)}]{deGroot}%
  \BibitemOpen
  \bibfield  {author} {\bibinfo {author} {\bibfnamefont {S.}~\bibnamefont
  {de~Groot}}\ and\ \bibinfo {author} {\bibfnamefont {W.}~\bibnamefont {van
  Leeuwen}},\ }\href@noop {} {\emph {\bibinfo {title} {Relativistic kinetic
  theory : principles and applications /}}}\ (\bibinfo  {publisher}
  {North-Holland},\ \bibinfo {year} {1980})\BibitemShut {NoStop}%
\bibitem [{\citenamefont {Huovinen}\ and\ \citenamefont
  {Molnar}(2009)}]{Huovinen:2008te}%
  \BibitemOpen
  \bibfield  {author} {\bibinfo {author} {\bibfnamefont {P.}~\bibnamefont
  {Huovinen}}\ and\ \bibinfo {author} {\bibfnamefont {D.}~\bibnamefont
  {Molnar}},\ }\href@noop {} {\bibfield  {journal} {\bibinfo  {journal}
  {Phys.Rev.}\ }\textbf {\bibinfo {volume} {C79}},\ \bibinfo {pages} {014906}
  (\bibinfo {year} {2009})}\BibitemShut {NoStop}%
\bibitem [{\citenamefont {El}\ \emph {et~al.}(2010)\citenamefont {El},
  \citenamefont {Muronga}, \citenamefont {Xu},\ and\ \citenamefont
  {Greiner}}]{El:2010mt}%
  \BibitemOpen
  \bibfield  {author} {\bibinfo {author} {\bibfnamefont {A.}~\bibnamefont
  {El}}, \bibinfo {author} {\bibfnamefont {A.}~\bibnamefont {Muronga}},
  \bibinfo {author} {\bibfnamefont {Z.}~\bibnamefont {Xu}}, \ and\ \bibinfo
  {author} {\bibfnamefont {C.}~\bibnamefont {Greiner}},\ }\href@noop {}
  {\bibfield  {journal} {\bibinfo  {journal} {Nucl. Phys.}\ }\textbf {\bibinfo
  {volume} {A848}},\ \bibinfo {pages} {428} (\bibinfo {year}
  {2010})}\BibitemShut {NoStop}%
\bibitem [{\citenamefont {{A.~El}}\ \emph {et~al.}()\citenamefont {{A.~El}},
  \citenamefont {Lauciello} \emph {et~al.}}]{Lauciello_paper}%
  \BibitemOpen
  \bibfield  {author} {\bibinfo {author} {\bibnamefont {{A.~El}}}, \bibinfo
  {author} {\bibfnamefont {F.}~\bibnamefont {Lauciello}},  \emph {et~al.},\
  }\href@noop {} {}\bibinfo {note} {Paper in preparation}\BibitemShut {NoStop}%
\bibitem [{\citenamefont {Gunion}\ and\ \citenamefont
  {Bertsch}(1982)}]{Gunion:1981qs}%
  \BibitemOpen
  \bibfield  {author} {\bibinfo {author} {\bibfnamefont {J.}~\bibnamefont
  {Gunion}}\ and\ \bibinfo {author} {\bibfnamefont {G.}~\bibnamefont
  {Bertsch}},\ }\href@noop {} {\bibfield  {journal} {\bibinfo  {journal}
  {Phys.Rev.}\ }\textbf {\bibinfo {volume} {D25}},\ \bibinfo {pages} {746}
  (\bibinfo {year} {1982})}\BibitemShut {NoStop}%
\bibitem [{\citenamefont {Migdal}(1956)}]{LPM}%
  \BibitemOpen
  \bibfield  {author} {\bibinfo {author} {\bibfnamefont {A.~B.}\ \bibnamefont
  {Migdal}},\ }\href@noop {} {\bibfield  {journal} {\bibinfo  {journal} {Phys.
  Rev.}\ }\textbf {\bibinfo {volume} {103}},\ \bibinfo {pages} {1811} (\bibinfo
  {year} {1956})}\BibitemShut {NoStop}%
\bibitem [{\citenamefont {Blaizot}\ \emph {et~al.}(1999)\citenamefont
  {Blaizot}, \citenamefont {Iancu},\ and\ \citenamefont
  {Rebhan}}]{PhysRevLett.83.2906}%
  \BibitemOpen
  \bibfield  {author} {\bibinfo {author} {\bibfnamefont {J.-P.}\ \bibnamefont
  {Blaizot}}, \bibinfo {author} {\bibfnamefont {E.}~\bibnamefont {Iancu}}, \
  and\ \bibinfo {author} {\bibfnamefont {A.}~\bibnamefont {Rebhan}},\
  }\href@noop {} {\bibfield  {journal} {\bibinfo  {journal} {Phys. Rev. Lett.}\
  }\textbf {\bibinfo {volume} {83}},\ \bibinfo {pages} {2906} (\bibinfo {year}
  {1999})}\BibitemShut {NoStop}%
\bibitem [{\citenamefont {Baym}\ \emph {et~al.}(1990)\citenamefont {Baym},
  \citenamefont {Monien}, \citenamefont {Pethick},\ and\ \citenamefont
  {Ravenhall}}]{PhysRevLett.64.1867}%
  \BibitemOpen
  \bibfield  {author} {\bibinfo {author} {\bibfnamefont {G.}~\bibnamefont
  {Baym}}, \bibinfo {author} {\bibfnamefont {H.}~\bibnamefont {Monien}},
  \bibinfo {author} {\bibfnamefont {C.~J.}\ \bibnamefont {Pethick}}, \ and\
  \bibinfo {author} {\bibfnamefont {D.~G.}\ \bibnamefont {Ravenhall}},\
  }\href@noop {} {\bibfield  {journal} {\bibinfo  {journal} {Phys. Rev. Lett.}\
  }\textbf {\bibinfo {volume} {64}},\ \bibinfo {pages} {1867} (\bibinfo {year}
  {1990})}\BibitemShut {NoStop}%
\bibitem [{\citenamefont {Arnold}\ \emph {et~al.}(2000)\citenamefont {Arnold},
  \citenamefont {Moore},\ and\ \citenamefont {Yaffe}}]{Arnold:2000dr}%
  \BibitemOpen
  \bibfield  {author} {\bibinfo {author} {\bibfnamefont {P.~B.}\ \bibnamefont
  {Arnold}}, \bibinfo {author} {\bibfnamefont {G.~D.}\ \bibnamefont {Moore}}, \
  and\ \bibinfo {author} {\bibfnamefont {L.~G.}\ \bibnamefont {Yaffe}},\
  }\href@noop {} {\bibfield  {journal} {\bibinfo  {journal} {JHEP}\ }\textbf
  {\bibinfo {volume} {11}},\ \bibinfo {pages} {001} (\bibinfo {year}
  {2000})}\BibitemShut {NoStop}%
\bibitem [{\citenamefont {Peshier}\ and\ \citenamefont
  {Cassing}(2005)}]{Peshier:2005pp}%
  \BibitemOpen
  \bibfield  {author} {\bibinfo {author} {\bibfnamefont {A.}~\bibnamefont
  {Peshier}}\ and\ \bibinfo {author} {\bibfnamefont {W.}~\bibnamefont
  {Cassing}},\ }\href@noop {} {\bibfield  {journal} {\bibinfo  {journal} {Phys.
  Rev. Lett.}\ }\textbf {\bibinfo {volume} {94}},\ \bibinfo {pages} {172301}
  (\bibinfo {year} {2005})}\BibitemShut {NoStop}%
\bibitem [{\citenamefont {Khvorostukhin}\ \emph {et~al.}(2010)\citenamefont
  {Khvorostukhin}, \citenamefont {Toneev},\ and\ \citenamefont
  {Voskresensky}}]{Khvorostukhin:2010cw}%
  \BibitemOpen
  \bibfield  {author} {\bibinfo {author} {\bibfnamefont {A.}~\bibnamefont
  {Khvorostukhin}}, \bibinfo {author} {\bibfnamefont {V.}~\bibnamefont
  {Toneev}}, \ and\ \bibinfo {author} {\bibfnamefont {D.}~\bibnamefont
  {Voskresensky}},\ }\href@noop {} {\  (\bibinfo {year} {2010})},\ \Eprint
  {http://arxiv.org/abs/1011.0839} {arXiv:1011.0839 [nucl-th]} \BibitemShut
  {NoStop}%
\bibitem [{\citenamefont {Plumari}\ \emph {et~al.}(2011)\citenamefont
  {Plumari}, \citenamefont {Alberico}, \citenamefont {Greco},\ and\
  \citenamefont {Ratti}}]{Plumari:2011mk}%
  \BibitemOpen
  \bibfield  {author} {\bibinfo {author} {\bibfnamefont {S.}~\bibnamefont
  {Plumari}}, \bibinfo {author} {\bibfnamefont {W.~M.}\ \bibnamefont
  {Alberico}}, \bibinfo {author} {\bibfnamefont {V.}~\bibnamefont {Greco}}, \
  and\ \bibinfo {author} {\bibfnamefont {C.}~\bibnamefont {Ratti}},\
  }\href@noop {} {\  (\bibinfo {year} {2011})},\ \Eprint
  {http://arxiv.org/abs/1103.5611} {arXiv:1103.5611 [hep-ph]} \BibitemShut
  {NoStop}%
\bibitem [{\citenamefont {Fochler}\ \emph {et~al.}(2010)\citenamefont
  {Fochler}, \citenamefont {Xu},\ and\ \citenamefont
  {Greiner}}]{Fochler:2010wn}%
  \BibitemOpen
  \bibfield  {author} {\bibinfo {author} {\bibfnamefont {O.}~\bibnamefont
  {Fochler}}, \bibinfo {author} {\bibfnamefont {Z.}~\bibnamefont {Xu}}, \ and\
  \bibinfo {author} {\bibfnamefont {C.}~\bibnamefont {Greiner}},\ }\href@noop
  {} {\bibfield  {journal} {\bibinfo  {journal} {Phys.Rev.}\ }\textbf {\bibinfo
  {volume} {C82}},\ \bibinfo {pages} {024907} (\bibinfo {year}
  {2010})}\BibitemShut {NoStop}%
\bibitem [{\citenamefont {Reining}\ \emph {et~al.}()\citenamefont {Reining}
  \emph {et~al.}}]{reining_paper}%
  \BibitemOpen
  \bibfield  {author} {\bibinfo {author} {\bibfnamefont {F.}~\bibnamefont
  {Reining}} \emph {et~al.},\ }\href@noop {} {\enquote {\bibinfo {title}
  {Extraction of shear viscosity in stationary states of relativistic particle
  systems},}\ }\bibinfo {note} {Paper in preperation}\BibitemShut {NoStop}%
\end{thebibliography}%

\end{document}